\documentclass[journal]{IEEEtran}
\usepackage{amssymb}
\usepackage{amsmath}
\usepackage{amsfonts}
\usepackage{algorithm}
\usepackage{algorithmic}
\usepackage{graphicx}

\makeatletter
\newtheorem{theorem}{Theorem}[section]

\newtheorem{definition}[theorem]{Definition}

\newtheorem{lemma}[theorem]{Lemma}

\newtheorem{remark}[theorem]{Remark}

\numberwithin{equation}{section}
\numberwithin{equation}{subsection}
\numberwithin{theorem}{subsection}

\newcommand{\C}{\mathbb{C}}
\newcommand{\eps}{\varepsilon}
\input{tcilatex}
\begin{document}

\title{Performance Estimates of the Pseudo-Random Method for Radar Detection
\smallskip }
\author{{\Large Alexander Fish and Shamgar Gurevich} \thanks{%
This material is based upon work supported by the Defense Advanced Research
Projects Agency (DARPA) award number N66001-13-1-4052. This work was also
supported in part by NSF Grant DMS-1101660 - "The Heisenberg--Weil
Symmetries, their Geometrization and Applications".\smallskip {}} \thanks{%
A. Fish is with School of Mathematics and Statistics, University of Sydney,
Sydney, NSW 2006, Australia. Email: alexander.fish@sydney.edu.au.} \thanks{%
S. Gurevich is with the Department of Mathematics, University of Wisconsin,
Madison, WI 53706, USA. Email: shamgar@math.wisc.edu.} }
\maketitle

\begin{abstract}
A performance of the pseudo-random method for the radar detection is
analyzed. The radar sends a pseudo-random sequence of length $N$, and
receives echo from $r$ targets. We assume the natural assumptions of
uniformity on the channel and of the square root cancellation on the noise.
Then for $r \leq N^{1-\delta}$, where $\delta > 0$, the following holds: (i) the probability of detection goes to one, 
and (ii)  the expected number of false targets goes to zero,  as $N$ goes to infinity.

\end{abstract}



\section{\textbf{Introduction\label{In}}}

\PARstart{A}{ radar} is designed to estimate the location and velocity of
objects in the surrounding space. The radar performs sensing by analyzing
correlations between sent and received (analog) signals. In this note we
describe the digital radar, i.e. we assume that the radar sends and receives
finite sequences. The reduction to digital setting can be carried out in
practice, see for example \cite{FGHSS}, and \cite{TV}.  \smallskip

Throughout this note we denote by $\mathbb{Z}_N$ the set of integers $%
\{0,1,...,N-1\}$ equipped with addition and multiplication modulo $N$. We
denote by $\mathcal{H}=%
\mathbb{C}
(%
\mathbb{Z}
_{N})$ the vector space of complex valued functions on $%
\mathbb{Z}
_{N}$ equipped with the standard inner product $\langle \cdot , \cdot \rangle
$, and refer to it as the \textit{Hilbert space of sequences}. We use the
notation $S_{\mathbb{C}}^{r-1}$ to denote the unit complex sphere in $%
\mathbb{C}^{r}$: 
\begin{equation*}
S_{\mathbb{C}}^{r-1} = \{ (z_1,\ldots,z_r) \in \mathbb{C}^r \, | \,
\sum_{k=1}^r |z_k|^2 = 1\}. 
\end{equation*}

\subsection{\textbf{Model of Digital Radar}\label{Model}}

We describe the discrete radar model which was derived in \cite{FGHSS}.  We
assume that a radar sends a sequence $S \in \mathcal{H}$ and receives as an
echo a sequence $R \in \mathcal{H}$.  The relationship between $S,$ and $R$
is given by the following equation:

\begin{equation}  \label{discr_channel}
R[n] = H(S)[n] + \mathcal{W}[n], \mbox{ } n \in \mathbb{Z}_N,
\end{equation}
where $H$, called the \textit{channel operator}, is defined by\footnote{%
We denote $e(t) = \exp(2\pi i t/N)$.} 
\begin{equation}  \label{operator}
H(S)[n] = \sum_{k=1}^{r} \alpha_k e(\omega_k n) S[n - \tau_k], \mbox{  } n
\in \mathbb{Z}_N,
\end{equation}
with $\alpha_k$'s the complex-valued attenuation coefficients associated
with target $k$, $\Vert \vec \alpha \Vert^2 = \sum_{k} | \alpha_k |^2 = 1$, $\tau_k \in \mathbb{Z}_N$
the time shift associated with target $k$, $\omega_k \in \mathbb{Z}_N$ the
frequency shift associated with target $k$, and $\mathcal{W}$ denotes a 
\textit{random} noise. The parameter  $r$ will be called the \textit{sparsity} of the channel. The
time-frequency shifts $(\tau_k,\omega_k)$ are related to the location and
velocity of target $k$. We denote the plane $\mathbb{Z}_N \times \mathbb{Z}_N
$ of all time-frequency shifts by $V$. We denote by $P$ the probability measure on the sample space generated by the random noise and attenuation coefficients.
\medskip

\begin{remark}
Let us elaborate on the constraint $\Vert \vec \alpha \Vert = 1$. In reality  $\Vert \vec \alpha \Vert \leq 1$.  However, we can rescale the received sequence $R$ to make $\Vert \vec \alpha \Vert = 1$. The  rescaling will not change the quality of the detection, as  evident from Section \ref{PRM}.
\end{remark}
\medskip

 We make the following assumption on
the distribution of $\mathcal{W}$: \medskip

\textit{Assumption} (\textbf{Square root cancellation}): For every $%
\varepsilon> 0$, there exists $c > 0$, such that for any $N^2$ vectors $%
u_1,\ldots,u_{N^2} \in S_\mathbb{C}^{N-1}$ we have 
\begin{equation*}
P\left(|\langle \mathcal{W},u_j \rangle | \leq N^{-\frac{1}{2} +
\varepsilon}, \, \,\, j=1,2,\ldots,N^2 \right) \geq 1 - e^{-c N}. 
\end{equation*}
\medskip

Note that an additive white gaussian noise (AWGN) of a constant, i.e.,
independent of $N$, signal-to-noise ratio (SNR) satisfies this assumption.
\smallskip

In addition, we make the following natural assumption on the distribution of the attenuation
coefficients $(\alpha_1,\ldots,\alpha_r)$ of the channel operator: \medskip

\textit{Assumption} (\textbf{Uniformity}): For any measurable subset $E \in
S_{\mathbb{C}}^{r-1}$ we have 
\begin{equation*}
P\left((\alpha_1,\ldots,\alpha_r) \in E \right) = \frac{Area(E)}{Area(S_{%
\mathbb{C}}^{r-1})}, 
\end{equation*}
where $Area$ denotes the unique up to scaling non-negative Borel measure on $S_{\C}^{r-1}$ which is invariant under all rotations of $\C^r$, i.e., elements in $SO_r(\C)$.
\medskip

The last natural assumption that we make is the independence of the noise and of the vector of attenuation coefficients of the channel operator:\medskip

\textit{Assumption} (\textbf{Independence}): The random sequences $\mathcal{W} \in \mathcal{H}$ and $\vec \alpha = (\alpha_1,\ldots,\alpha_r) \in S_{\mathbb{C}}^{r-1}$ are independent. 

\subsection{\textbf{Objectives of the Paper}}

The main task of the digital radar system is to extract the channel
parameters $(\tau _{k},\omega _{k})$, $k=1,...,r,$ using $S$ and $R$
satisfying (\ref{discr_channel}). One of the most popular methods for
channel estimation is the pseudo-random (PR) method. In this note we
describe the PR method and analyze its performance.

\section{\textbf{Ambiguity Function and Pseudo-Random Method}}

A classical method to estimate the channel parameters in (\ref{discr_channel}%
) is the \textit{pseudo-random method }\cite{GG, GHS, HCM, TV, V}. It uses
two ingredients - the ambiguity function, and a pseudo-random
sequence.\smallskip

\subsection{\textbf{Ambiguity Function}}

In order to reduce the noise component in (\ref{discr_channel}), it is
common to use the ambiguity function that we are going to describe now. We
consider the time-frequency shift operators $\pi (\tau ,\omega ),$ $\tau
,\omega \in 
\mathbb{Z}
_{N},$ which act on $f\in \mathcal{H}$ by%
\begin{equation}
\left[ \pi (\tau ,\omega )f\right] [n]= e(\omega n)\cdot f[n-\tau ], \mbox{ }
n \in \mathbb{Z}_N  \label{HO}
\end{equation}%
The \textit{ambiguity function }of two sequences $f,g\in \mathcal{H} $ is
defined as the $N\times N$ matrix 
\begin{equation}
\mathcal{A(}f,g)[\tau ,\omega ]=\left\langle \pi (\tau ,\omega
)f,g\right\rangle ,\text{ \ }\tau ,\omega \in 
\mathbb{Z}
_{N}.  \label{AF}
\end{equation}
\smallskip

\begin{remark}[\textbf{Fast Computation of Ambiguity Function}]
\label{FC}The restriction of the ambiguity function to a line in the
time-frequency plane, can be computed in $O(N\log N)$ arithmetic operations
using fast Fourier transform. For more details, including explicit
formulas---see Section V of \cite{FGHSS}. Overall, we can compute the entire
ambiguity function in $O(N^{2}\log N)$ operations.\smallskip
\end{remark}


\subsection{\textbf{Pseudo-Random Sequences}}

\begin{figure}[ht]
\includegraphics[clip,height=4cm]{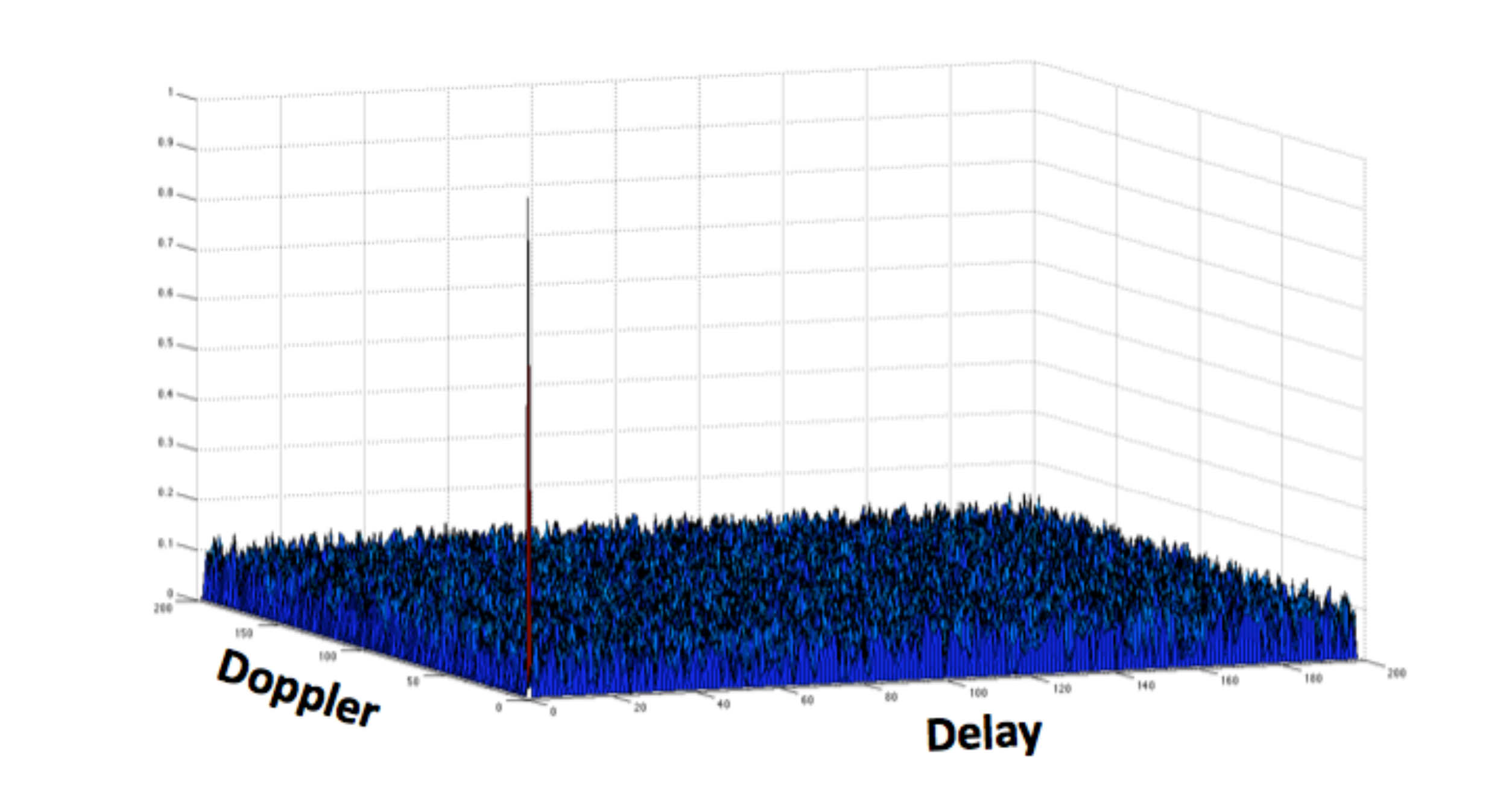}\newline
\caption{Profile of $\mathcal{A}(\protect\varphi ,\protect\varphi )$ for $%
\protect\varphi $ pseudo-random sequence.}
\label{PRFigure}
\end{figure}

 We say that a norm-one sequence $\varphi \in \mathcal{H}$ is $B $-%
\textit{pseudo-random, }$B\in 
\mathbb{R}
$\textit{---}see Figure \ref{PRFigure} for illustration---if for every $%
\left( \tau ,\omega \right) \neq (0,0)$ we have \textit{\ }%
\begin{equation}
\left\vert \mathcal{A}(\varphi ,\varphi )[\tau ,\omega ]\right\vert \leq B/%
\sqrt{N}.  \label{pr}
\end{equation}%
There are several constructions of families of pseudo-random (PR) sequences
in the literature---see \cite{GG, GHS} and references therein.

\subsection{\textbf{Pseudo-Random (PR) Method}}
\label{PRM}

Consider a pseudo-random sequence $\varphi $, and assume for simplicity that 
$B=1$ in (\ref{pr}). Then we have 
\begin{eqnarray}
&&\mathcal{A}(\varphi ,H(\varphi ))[\tau ,\omega ]  \label{prm} \\
&=&\left\{ 
\begin{array}{c}
\widetilde{\alpha }_{k}+\tsum\limits_{j\neq k}\widetilde{\alpha }_{j}/\sqrt{N%
},\text{ \ if }\left( \tau ,\omega \right) =\left( \tau _{k},\omega
_{k}\right) ,\text{ }1\leq k\leq r; \\ 
\tsum\limits_{j}\widehat{\alpha }_{j}/\sqrt{N},\text{ \ \ \ \ \ otherwise, \
\ \ \ \ \ \ \ \ \ \ \ \ \ \ \ \ \ \ \ \ \ \ \ \ \ }%
\end{array}%
\right.  \notag
\end{eqnarray}%
where $\widetilde{\alpha }_{j},$ $\widehat{\alpha }_{j},$ $1\leq j\leq r,$
are certain multiples of the $\alpha _{j}$'s by complex numbers of absolute
value less or equal to one. In particular, we can compute the time-frequency
parameter $\left( \tau _{k},\omega _{k}\right) $ if the associated
attenuation coefficient $\alpha _{k}$ is sufficiently large, i.e., it appears as a
``peak" of $\mathcal{A}(\varphi ,H(\varphi )).$ 
\medskip

\begin{definition}[\textbf{$\delta$-peak}]
Let $\delta > 0$. We say that at $v \in V$ the ambiguity function of $f$ and $g$ has $\delta$-peak, if 
$$\left| \mathcal{A}(f,g)[v] \right| \geq  N^{-1/2 + \delta}.$$
\end{definition}
\medskip

\begin{figure}[ht]
\includegraphics[clip,height=4cm]{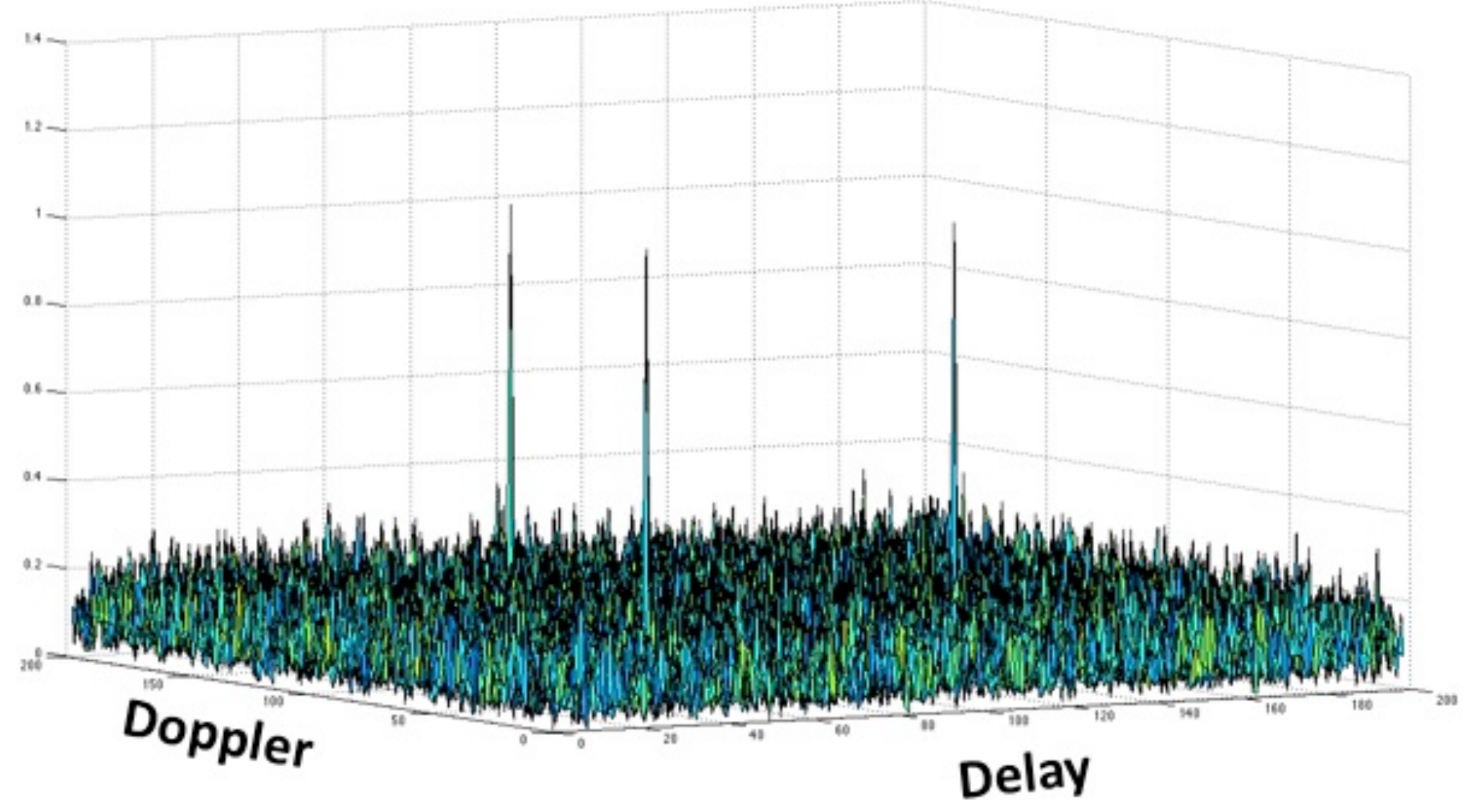}\newline
\caption{Profile of $\mathcal{A}(S,R )$ for a pseudo-random sequence $S$,
and a channel of sparsity three.}
\label{three-peaks}
\end{figure}

Below we describe---see
Figure \ref{three-peaks}---the PR method.

\begin{algorithm}
\underline{\textbf{Pseudo-Random Method}}\smallskip 

\begin{description}

\item[\textbf{Input: }] $\,$
Pseudo-random sequence $S \in \mathcal{H}$, the echo $R$, and a parameter  $\delta > 0$.

\smallskip 

\item[\textbf{Output: }] $\,\,\,$ Channel parameters.

\end{description}

\smallskip

 Compute $\mathcal{A}(S,R)$ on $V$ and return those time-frequency shifts at which  the $\delta$-peaks occur. 
\end{algorithm}
%
\medskip

We  call the above computational scheme  the \textbf{PR method with parameter $\delta$}.
\medskip

Notice that the arithmetic complexity of the PR method is $O(N^{2}\log N),$
using Remark \ref{FC}.

\section{\textbf{Performance of the PR Method}}

First, we introduce two important  quantities that measure the performance of a
detection scheme. These are the probability of detection and the expected number of false targets.
Then, we formulate the main result of this note which provides a
quantitative statement about the performance of the PR method.

Assume that we have $r$ targets out of the $N^2$ possible---see Section \ref%
{Model}. Also, assume that some random data is associated with these targets
and influencing the performance of a detection scheme. For example, in our
setting the random data consists of (i)  the attenuation coefficients associated with the
targets, and (ii) the noise. In general, we model the randomness of the data by associating a
probability space $(\boldsymbol{\Omega},P)$ to the set of targets. For every 
$\boldsymbol{\omega}$ in the space $\boldsymbol{\Omega}$ we denote by $\mathcal{N}_t(%
\boldsymbol{\omega})$ and $\mathcal{N}_f(\boldsymbol{\omega})$, the number of true and
false targets detected by the  scheme, respectively.  We define the \textit{%
probability of detection} by 
\begin{equation*}
P_D = P(\mbox{a true target is detected}) = \frac{1}{r} \int_{\boldsymbol{%
\Omega}} \mathcal{N}_t(\boldsymbol{\omega}) dP(\boldsymbol{\omega}), 
\end{equation*}
and the \textit{expected number of false targets}   by 
\begin{equation*}
E_{FT} = E(\mathcal{N}_f)
= \int_{\boldsymbol{\Omega}} \mathcal{N}_f(\boldsymbol{\omega}%
) dP(\boldsymbol{\omega}). 
\end{equation*}
The main result of this note is the following: \medskip

\begin{theorem}[\textbf{Performance of PR method}]
\label{main_thm} Assume  the channel operator (\ref{operator})
satisfies the uniformity,  square root cancellation,  and  independence assumptions. Then for $r
\leq N^{1-\delta}$, where $\delta > 0$, we have   $P_D \to 1$, and $E_{FT} \to 0
$, as $N \to \infty$, for the PR method with  parameter $\delta/4$.%
\end{theorem}
\medskip

\begin{remark}[\textbf{Rate of convergence}] 
We suspect that the true rate of convergence in  $P_D \to 1$, and $E_{FT} \to 0$, as $N \to \infty$, is polynomial. In fact, our proof of Theorem \ref{main_thm} confirms  that the  rate of convergence   is at least polynomial. 
\end{remark}

\section{\textbf{Conclusions\label{Co}}}

The obtained estimates in Theorem \ref{main_thm} show that the PR method is
effective in terms of   performance of  detection,  in the regime  $r \leq N^{1-\delta}$, for $\delta > 0$. We would like to note that if $r \geq N^{1+\eps}$, for  $\eps > 0$,  then the performance of the PR method deteriorates 
since the noise influence becomes dominant.  \medskip

\section{\textbf{Proof of Theorem} \protect\ref{main_thm}}

Before giving a formal proof, we provide a sketch. To detect the parameters
of the channel operator $H$, the PR method evaluates the ambiguity function
of $R$, and $S$ at $v_k$: 
\begin{equation*}
\mathcal{A}(S,R)(v_k) = \alpha_k + \sum_{ j \neq k} \alpha_j \langle \pi(v_k)
S, \pi(v_j) S \rangle + \langle \pi(v_k) S, \mathcal{W} \rangle. 
\end{equation*}
Let us denote by $c_k = \sum_{ j \neq k} \alpha_j \langle \pi(v_k) S,
\pi(v_j) S \rangle$ the $k$-th cross term, and by $\nu_k = \langle \pi(v_k)
S, \mathcal{W} \rangle$ the $k$-th noise component. Then we have 
\begin{equation}
\mathcal{A}(S,R)(v_k) = \alpha_k + c_k + \nu_k.
\end{equation}
The parameter $v_k$ is detectable by the PR method if the main term $\alpha_k$
is much larger than the $k$-th cross term $c_k$, and the noise component $%
\nu_k$. For a random point on the unit sphere $S_{\mathbb{C}}^{r-1}$, by
``concentration" most of its coordinates are of absolute value approximately
equal to $1/\sqrt{r}$. Thus, if $r \leq N$, the magnitude of most of $\alpha_k$'s  is greater than $1/\sqrt{N}$. Another instance of the concentration
phenomenon guarantees that for most channels, the magnitude of the cross term $c_k$ is
smaller than $1/\sqrt{N}$. Finally, the square root cancellation assumption
on the noise guarantees that the magnitude of the noise term $\nu_k$ is much smaller than $1/\sqrt{N}$. \smallskip

We begin the formal proof of the theorem with auxiliary lemmata. In Section %
\ref{Proofs_Section} we prove these statements. \smallskip

\begin{lemma}[\textbf{Largeness of a slice}]
\label{large_coord} Let $(\alpha_1,\ldots,\alpha_r)$ be a uniformly chosen
point on $S_{\mathbb{C}}^{r-1}$, and fix $k \in \{1,\ldots,r\}$. There exists $K
> 0$ (independent of $r$ and $k$), such that for every $\varepsilon > 0$ we
have 
\begin{equation*}
P\left(|\alpha_k| \geq \varepsilon \right) \geq 1 - K \sqrt{r}
\varepsilon. 
\end{equation*}
\end{lemma}

\smallskip

\begin{lemma}[\textbf{Intersectivity}]
\label{prob_lem} Let $E_1,\ldots,E_r$ be events in a probability space $(%
\boldsymbol{\Omega},P)$, such that $P(E_k) \geq 1 - r^{-\delta}$, $%
k=1,\ldots,r$, for some $\delta > 0$. Then for the event 
\begin{equation*}
E = \{ \boldsymbol{\omega} \in \boldsymbol{\Omega} \, \, | \, \, \boldsymbol{%
\omega} \mbox{ is in at least } n(r,\delta) \mbox{ of } E_k\mbox{'s} \}, 
\end{equation*}
where 
\begin{equation*}
n(r,\delta) = \lfloor (1-r^{-\delta/2}) r \rfloor, 
\end{equation*}
we have 
\begin{equation*}
P(E) \geq 1 - r^{-\delta/2}. 
\end{equation*}
\end{lemma}

\smallskip

\begin{lemma}[\textbf{Almost orthogonality}]
\label{mixed_terms} Let $\ell > 0$, and let $\vec{z_j} =
(z_1^j,\ldots,z_r^j) \in \mathbb{C}^r$, $j=1,\ldots,r^{\ell}$, be vectors
satisfying 
\begin{equation*}
\sum_{k=1}^r |z_k^j|^2 \leq C^2 \frac{r}{N}, \mbox{ for some } C > 0. 
\end{equation*}
Then for any $\delta > 0$, there exists $\beta > 0$, such that for a
uniformly chosen random point $\vec{\alpha} \in S_{\mathbb{C}}^{r-1}$ we
have 
\begin{equation*}
P\left( \bigcap_{j=1}^{r^{\ell}} \left\{\left| \langle \vec{\alpha}, \vec{z_j}\rangle \right| \leq \frac{%
C r^{\delta}}{\sqrt{N}} \right\} \right) \geq 1 - e^{-\beta r^{2 \delta}}. 
\end{equation*}
\end{lemma}

\medskip

\noindent

\begin{proof}[\textbf{Proof of Theorem \protect\ref{main_thm}.}]
Assume that $r \leq N^{1-\delta}$, for some $\delta > 0$. Let $R = HS + 
\mathcal{W}$, where $S$ is a ($B=1$)-pseudo-random sequence in $\mathcal{H}$%
, $H$ is a channel of sparsity $r$ with uniformly distributed attenuation
coefficients given by (\ref{operator}), and  $\mathcal{W}$ satisfies the
square root cancellation assumption. We denote $v_k = (\tau_k, \omega_k), k
=1,\ldots,r$, and assume that at the receiver we perform PR method with parameter $\delta/4$. \smallskip

\textbf{(A)} Proof of ``$P_D \to 1$ as $N \to \infty$". \medskip

We consider two cases. \smallskip

\textbf{Case 1.} $r \geq \log{N}$. \smallskip

Denote by $E_k = \{\boldsymbol{\omega} \in \boldsymbol{\Omega} \,\, | \,\, |
\alpha_k(\boldsymbol{\omega})| \geq N^{-1/2+\delta/3}\}$ for $k=1,\ldots,r$.
Since $r \leq N^{1-\delta}$ by Lemma \ref{large_coord} there exists $K > 0$ such that we have

\begin{equation*}
P(E_k) \geq 1 - Kr^{-\frac{ \delta/6}{1 - \delta}}. 
\end{equation*}
Therefore, for sufficiently large $N$, we have 
\[
P(E_k) \geq 1 - r^{-\frac{ \delta/7}{1 - \delta}}.
\]
Denote by 
\begin{equation*}
E = \{\boldsymbol{\omega} \in \boldsymbol{\Omega} \, \, | \,\, \boldsymbol{%
\omega} \mbox{ is in at least } (1 - r^{-\frac{ \delta}{14(1-\delta)}}) r %
\mbox{ of } E_k \mbox{'s}\}. 
\end{equation*}
By Lemma \ref{prob_lem} we have 
\begin{equation}  \label{eq_1}
P(E) \geq 1 - r^{-\frac{\delta}{14(1 - \delta)}}.
\end{equation}
Since $r \leq N^{1-\delta}$, we have $r^{\frac{\delta/5}{1-\delta}} \leq N^{\delta/5}$. Therefore, by Lemma
\ref{mixed_terms}, there exists $\beta > 0$ such that 
\[
P\left(\bigcap_{k=1}^r \left\{\left| \sum_{j \neq k} \alpha_j \langle \pi(v_k) S, \pi(v_j)S \rangle
\right| \leq N^{-\frac{1}{2} + \frac{\delta}{5}} \right\}\right) 
\]

\begin{equation}  \label{eq_2}
\geq 1 - e^{-\beta r^{\frac{2\delta}{5(1 - \delta)}}}.
\end{equation}

It follows from (\ref{prm}), (\ref{eq_1}), (\ref{eq_2}),  the square root
cancellation and independence assumptions, that with probability greater or equal
than $1 - r^{-\frac{\delta}{15(1-\delta)}}$, at least $(1 - r^{-\frac{%
\delta}{14(1 - \delta)}})r$ of the channel parameters of $H$ are detectable.

The latter implies that $P_D \geq (1 - r^{-\frac{\delta}{15(1 - \delta)}}) (1
- r^{-\frac{\delta}{14(1 - \delta)}})$ for $N$ sufficiently large. Therefore
we have $P_D \to 1$ as $N \to \infty$. \bigskip

\textbf{Case 2.} $r \leq \log{N}$. \smallskip

By Lemma \ref{large_coord}, there exists $K > 0$ such that 
\begin{equation*}
P\left(\bigcap_{k=1}^{r} \left\{ |\alpha_k| \geq N^{-1/2 + \delta/3} \right\}\right) \geq 1 - K (\log{N})^{3/2}
N^{-1/2 + \delta/3}. 
\end{equation*}
By Cauchy-Schwartz inequality we have for all $k=1,\ldots,r$: 
\begin{equation*}
\left| \sum_{j \neq k} \alpha_j \langle \pi(v_k) S, \pi(v_j) S\rangle\right|
\leq \frac{\sqrt{r}}{\sqrt{N}}  \leq \sqrt{\frac{\log{N}}{N}}. 
\end{equation*}
It follows from (\ref{prm}), square root cancellation assumption on the
noise, and the last two inequalities that 
for sufficiently large $N$, all $r$ channel parameters of the operator $H$
are detectable with probability greater or equal than $1 - N^{-1/2 + \delta/2}$.
Therefore, for sufficiently large $N$ we have 
\begin{equation*}
P_D \geq 1 -  N^{-1/2 + \delta/2}. 
\end{equation*}
The latter implies that $P_D \to 1$ as $N \to \infty$. \bigskip

\textbf{(B)} Proof of ``$E_{FT} \to 0$ as $N \to \infty$". \medskip

\textbf{Case 1.} $r \geq N^{\delta/3}$.

By Lemma \ref{mixed_terms}, there exists $\beta > 0$ such that we have
\begin{equation*}
P\left(\bigcap_{v \not \in supp(H) } \left\{ \left| \sum_{k=1}^r \alpha_k \langle \pi(v) S, \pi(v_k)S \rangle \right|
\leq N^{-\frac{1}{2} + \frac{\delta}{5}} \right\} \right) 
\end{equation*}
\[
\geq  1 - e^{-\beta r^{\frac{2\delta}{5(1 - \delta)}}},
\]
where $supp(H) \subset V$ is the set of all channel parameters of $H$.
It follows from (\ref{prm}), the square root cancellation and independence assumptions, and the last inequality that with probability greater or equal than $%
1 - e^{-\frac{\beta}{2} r^{\frac{2\delta}{5(1 - \delta)}}}$ the PR method will not detect any wrong channel
parameters. The latter implies that $E_{FT} \to 0$ as $N \to \infty$.

\textbf{Case 2.} $r \leq N^{\delta/3}$.

By Cauchy-Shwartz inequality we have that for every $v \not \in supp(H)$: 
\begin{equation*}
\left| \sum_{k=1}^r \alpha_k \langle \pi(v) S, \pi(v_k) S\rangle\right| \leq 
\frac{\sqrt{r}}{\sqrt{N}} \leq N^{-1/2 + \delta/6}. 
\end{equation*}
It follows from (\ref{prm}), the square root cancellation assumption on the
noise, and the last inequality that there exists $c > 0$ such that for $N$
sufficiently large we have 
\begin{equation*}
P\left( \bigcap_{v \not \in
supp(H)} \left\{\left|  \langle \pi(v)S, R \rangle \right| < N^{-1/2 + \delta/5}\right\}  \right) \geq 1 - e^{-cN}. 
\end{equation*}
 The
latter implies that the PR method with parameter $\delta/4$
satisfies $E_{FT} \to 0$ as $N \to \infty$.
\end{proof}

\section{\textbf{Proofs of Lemmata\label{Proofs_Section}}}

\begin{proof}[\textbf{Proof of Lemma \protect\ref{large_coord}}]
We identify the Borel probability space on $S_{\mathbb{C}}^{r-1}$ invariant
under all rotations with the Borel probability space $S^{2r-1}$ of the real
unit sphere invariant under all rotations. Recall that 
\begin{equation*}
S^{2r-1} = \{ (x_1,y_1,x_2,y_2,\ldots,x_r,y_r) \in \mathbb{R}^{2r}\,\, |
\,\, \sum_{k=1}^{r} x_k^2 + y_k^2 =1 \}. 
\end{equation*}
Without loss of generality, it is enough to prove the statement for $k = 1$.
Let $\alpha_1 = x_1 + i \cdot y_1$. We use the notation $(\boldsymbol{\Omega}%
,P)$ for the probability space on $S^{2r-1}$ invariant under all rotations
in $\mathbb{R}^{2r}$.
For any $\rho > 0$, and any dimension $n$, we denote the real sphere of
radius $\rho$ in $\mathbb{R}^n$ by $S_{\rho}^{n-1}$: 
\begin{equation*}
S_{\rho}^{n-1} = \{(t_1,\ldots,t_n) \in \mathbb{R}^n \,\, | \,\,
\sum_{k=1}^n t_k^2 = \rho^2\}. 
\end{equation*}
Let $\eps > 0$. By Fubini's theorem and using the homogeneity of the Lebesgue measure we get 
\begin{equation*}
P(\boldsymbol{\omega} \in \boldsymbol{\Omega} \,\, | \,\,|x_1(\boldsymbol{%
\omega})| \leq \varepsilon ) 
\end{equation*}
\begin{equation*}
= \frac{2}{Area(S_{1}^{2r-1})} \cdot
\int_0^{\varepsilon} Area(S_{\sqrt{1-t^2}}^{2r-2})dt 
\end{equation*}
\begin{equation*}
= 2 \frac{Area(S_1^{2r-2})}{Area(S_1^{2r-1})} \int_0^{\varepsilon} (1-t^2)^{%
\frac{2r-2}{2}} dt. 
\end{equation*}
It is well known that 
\begin{equation*}
\frac{Area(S_1^{n-1})}{Area(S_1^{n})} \longrightarrow_{n \to \infty} \sqrt{%
\frac{n}{2 \pi}}. 
\end{equation*}
Therefore, for $r$ large enough we have 
\begin{equation*}
P(\boldsymbol{\omega} \in \boldsymbol{\Omega} \,\, | \,\,|x_1(\boldsymbol{%
\omega})| \leq \varepsilon ) \leq \sqrt{\frac{2(2r-1)}{\pi}} \int_0^{\varepsilon}
(1-t^2)^{r-1} dt 
\end{equation*}
\[
\leq \sqrt{\frac{4r}{\pi}} \cdot \varepsilon. 
\]
Finally, the containment $\{ \boldsymbol{\omega} \in \boldsymbol{\Omega}
\,\, | \,\, |\alpha_1(\boldsymbol{\omega})| \leq \varepsilon\} \subset \{ 
\boldsymbol{\omega} \in \boldsymbol{\Omega} \,\, | \,\, |x_1(\boldsymbol{%
\omega})| \leq \varepsilon\} $ together with last inequality imply the
statement of the lemma.
\end{proof}

\medskip

\begin{proof}[\textbf{Proof of Lemma \protect\ref{prob_lem}}]
Denote by $\gamma = P(E)$, and by $f_k = \chi_{E_k}$, $k=1,\ldots,r$. Then
we have 
\begin{equation*}
\int \sum_{k=1}^{r} f_k dP \geq r \cdot (1-r^{-\delta}). 
\end{equation*}
On the other hand, we have 
\begin{equation*}
\int \sum_{k=1}^{r} f_k dP = \int_E \sum_{k=1}^{r} f_k dP + \int_{E^c}
\sum_{k=1}^{r} f_k dP 
\end{equation*}
\begin{equation*}
\leq \gamma \cdot r + (1 - \gamma) \cdot (1 - r^{-\delta/2})\cdot r. 
\end{equation*}
The last inequality implies that 
\begin{equation*}
1 - r^{-\delta} \leq \gamma \cdot r^{-\delta/2} + (1 - r^{-\delta/2}). 
\end{equation*}
It implies 
\begin{equation*}
\gamma \geq 1 - r^{-\delta/2}. 
\end{equation*}
\end{proof}

\medskip

\begin{proof}[\textbf{Proof of Lemma \protect\ref{mixed_terms}}]
Let $\eps > 0$. We proceed similarly to the proof of Lemma \ref{large_coord}. Since
\[
\{ \boldsymbol{\omega} \in \boldsymbol{\Omega} \, | \, |\alpha_1(\boldsymbol{\omega})| > \eps \} 
\]
\[
\subset 
 \{ \boldsymbol{\omega} \in \boldsymbol{\Omega} \, | \, |x_1(\boldsymbol{\omega})| > \eps/2\} \cup \{ \boldsymbol{\omega} \in \boldsymbol{\Omega} \, | \, |y_1(\boldsymbol{\omega})| > \eps/2\}, 
\]
and
\[
P(|x_1| > \eps/2) = P (|y_1| > \eps/2) = 2 \frac{Area(S_1^{2r-2})}{Area(S_1^{2r-1})} \int_{\eps/2}^1 (1-t^2)^{r-1} \,dt,
\]
we conclude that there exists $K > 0$ such that 
\[
P(|\alpha_1| > \eps) \leq K \sqrt{r} \int_{\eps/2}^{1} (1-t^2)^{r-1} \, dt.
\]
If $\eps = r^{-1/2+\delta}$, then there exists $\beta' > 0$ such that
 $$(1 - (\eps/2)^2)^{r-1} \leq e^{-\beta' r^{2\delta}}.$$
  This implies that there exists $\beta'' > 0$ such that 
\[
P(|\alpha_1| \geq r^{-1/2+ \delta}) \leq e^{-\beta'' r^{2 \delta}}.
\] 
 By the rotation invariance of the
Lebesgue measure on $S^{2r-1}$, it follows that there exists $\beta > 0$ such that for any set of directions $%
\vec{\theta_1},\ldots,\vec{\theta_{r^{\ell}}} \in S^{2r-1}$ we have 
\begin{equation*}
P\left(\bigcup_{j=1}^{r^{\ell}} \left\{ \boldsymbol{\omega} \in \boldsymbol{\Omega} \, | \, |\langle \vec \alpha(\boldsymbol{\omega}), \vec{\theta_j}\rangle| \geq r^{-1/2 + \delta}\right\}\right) 
\end{equation*}
\begin{equation*}
\leq
r^{\ell} \cdot e^{-\beta'' r^{2\delta}} \leq e^{-\beta r^{2\delta}}.
 \end{equation*}
The latter implies the statement of the lemma.
\end{proof}

\medskip


\textbf{Acknowledgements. }We are grateful to our collaborators A. Sayeed, K. Scheim
and O. Schwartz, for many discussions related to the research reported in
these notes. Also, we thank anonymous referees for numerous suggestions.
And, finally, we are grateful to G. Dasarathy who kindly agreed to present this paper at the conference.

\end{document}